\documentclass[aps,unsortedaddress,a4paper,amsmath,amssymb,floatfix,showpacs]{revtex4-1}

\usepackage[latin1]{inputenc}
\usepackage{graphicx}
\usepackage{dcolumn}
\usepackage{bm}
\usepackage{times}
\usepackage{color}
\usepackage{float}
\usepackage{booktabs}

\newcommand\varpm{\mathbin{\vcenter{\hbox{%
	\oalign{\hfil$\scriptstyle+$\hfil\cr
	\noalign{\kern-.3ex}$\scriptscriptstyle({-})$\cr}%
}}}}

\begin{document}

\title{Thermoelectric response enhanced by surface/edge states in physical nanogaps}
%\title{Enhanced thermoelectric properties in physical nanogaps}

\author{V\'{\i}ctor M. Garc\'{\i}a-Su\'arez}\email{vm.garcia@cinn.es}
\affiliation{Department of Physics, University of Oviedo \& CINN, 33007 Oviedo, Spain}

\date{\today}

\begin{abstract}
Current solid-state thermoelectric converters have rather poor performances, which usually render them useless for practical applications. This problem is evidenced by the small figures of merit of typical thermoelectric materials, which tend to be much smaller than 1. Increasing this parameter is then key for the development of functional devices in technologically viable applications that can work optimally. We propose here a feasible and effective design of new thermoelectric systems based on physical gaps in nanoscale junctions. We show that, depending on the type of features, i.e. the character of surface/edge states, on both sides of the gap, it is possible to achieve rather high figures of merit. In particular, we show that for configurations that have localized states at the surfaces/edges, which translate into sharp resonances in the transmission, it is possible to achieve large Seebeck coefficients and figures of merit by carefully tuning their energy and their coupling to other states. We calculate the thermoelectric coefficients as a function of different parameters and find non-obvious behaviours, like the existence of a certain coupling between the localized and bulk states for which these quantities have a maximum. The highest Seebeck coefficients and figures of merit are achieved for symmetric junctions, which have the same coupling between the localized state and the bulk states on both sides of the gap. The features and trends of the thermoelectric properties and their change with various parameters that we find here cannot only be applied to systems with nanogaps, but also to many other nanoscale junctions, such as e.g. those that have surface states or states localized near the contacts between the nanoscale object and the electrodes. The model presented here can therefore be used to characterize and predict the thermoelectric properties of many different nanoscale junctions and can also serve as a guide for studying other systems. These results pave then the way for the design and fabrication of stable next-generation thermoelectric devices with robust features and improved performances.
\end{abstract}

\pacs{72.20.Pa,73.63.?b,85.35.Ds,71.15.?m}

\maketitle

\section{Introduction}

Solid state thermoelectric converters are key to designing systems that involve both the conversion electricity to heat in charge-driven cooling systems and heat to electricity in heat-driven current generators. These systems are essential for different industrial and technological processes, which range from applications in the automotive industry to space applications, biotechnological applications,  and waste of heat in industrial processes, to name a few \cite{Dis99,Sal02,Tri06,Bel08,Sny08,Sol19,Lei20,Fin21}. Improving current thermoelectric converters and designing new ones with enhanced capabilities is expected then to have a substantial impact in several scientific and technological areas and shall help to achieve different objectives related to energy saving and transformation, green energies and zero-carbon emissions \cite{Bel08}.

Current solid state thermoelectric converters can already be used in specialized applications such as spacecrafts \cite{Maz19}, but many of them present, however, a series of problems which reduce their performance and make them impractical for some technological applications. Among these problems stand specially out the small efficiency and poor performance of many of such systems integrated in different devices. For instance, the largest figures of merit in present-day solid state converters are of the order of 1 \cite{SuhBo,Hsu04,Dre07,Her08,Ola17}, with the highest ones achieved until now of around 3 \cite{Zha14,Cha18,Zho21}. These values can be reached mainly for bulk materials with nanostructured lattices. However, such values are clearly not enough to achieve efficient thermoelectric conversion nor fabricate very efficient and effective converters that could be employed in applications for real-life systems.

The main issues that severely affect the thermoelectric performance and lead to very poor performances come, on one hand, from the relatively small Seebeck coefficients, and, on the other hand, from the relatively large phonon conductances of most solid-state thermoelectric materials. The first factor is key to transform a temperature gradient into a potential difference by means of the Seebeck effect (or the other way round, through the Peltier effect) and the second factor, which is added to the electronic thermal conductance \cite{Mur08}, is key to ensure, when it is small enough, that the temperature gradient is kept and does not vanish, so that the thermal induced voltage difference or electrical current is maintained. However, the relatively small Seebeck coefficients and large phonon conductances of most solid state converters render them useless or unable to work properly, with a rather poor performance compared to other devices used in the recent past. 

In order to solve the problems that severely affect current devices and hamper their efficiency, various approaches that tackle different physical aspects have been proposed and implemented over the past years. These approaches include, for instance, the increase of the scattering of phonons in materials that include different sublattices or the nanostructuration of materials in different ways \cite{BhaBo,SlaBo}. This last approach in particular, i.e. nanostructuring certain parts or a whole material by including nanometric defects, has proved to be promising for designing new materials with rather high Seebeck coefficients ($S$) and figures of merit ($ZT$). Some examples include the use of nanoscale junctions, such as molecular electronic systems, with rather high $ZT$ \cite{Fin09}, or two dimensional materials such as graphene (see e.g. \cite{Dol15,Har18,Ahm22} and references therein). However, such systems are rather difficult to fabricate in practice and their relatively high figures of merit can only be reached within a very narrow and unstable window of parameters. Besides, in most of the calculations used to model such materials and predict their thermoelectric properties, the phonon thermal conductance, which is an important limiting factor that can severely reduce the size of $ZT$, is not taken into account and, therefore, it is not clear in many cases how efficiently these systems can work as thermoelectric converters.

In this work we propose the development of efficient nanoscale thermoelectric systems based on physical gaps between surfaces of bulk materials or 2D layers. We stress that such systems are real and some of them, such as those based on graphene layers, have already been characterized to some extent experimentally and simulated with ab-initio techniques \cite{GarNs18}. The proposed model describes then a specific type of physical systems and studies the main parameters that affect their thermoelectric performance. We describe and model the properties of these systems and show under which conditions thy display the highest performance. Such systems include, as commented before, graphene gaps, which can be fabricated with electroburning techniques \cite{PriNL11,NefNl14,GehNL16,Abb17} or with mechanically controllable break junctions \cite{CanNN18} and are currently being profusely studied by several groups worldwide as possible platforms to design new molecular electronics and nanoscale components. Examples of these last devices comprise the coupling of molecular wires, crossover molecules or other similar systems between graphene electrodes. An interesting property of such nanoscale devices is, however, that, even without any bridging element, they are expected to be reliable and display in some cases a rather good thermoelectric response generated, on one hand by the sharp transmission produced by different features on the edges of both sides of the gap \cite{Gar20}, as we shall see, and, on the other hand, by the lack of phonon transport across physically separated electrodes. The first factor, which is typical for instance of systems such as heavily doped semiconductors \cite{Lin00}, is key to generate large Seebeck coefficients, while the second one leads to the elimination of an important limiting effect in most solid-state thermoelectric materials. We note, however, in some cases phonons can still cross the gap and give sizeable thermal conductances \cite{Pru10}. This has also studied recently in different systems and it was found in many cases that such conductances cannot be ignored \cite{Sel12,Fon19,Xio20,Alk20,Tok22,Guo22}. We will therefore use an approximation to calculate such contribution (see below).

We show then that, in stable designs and under some conditions, it is possible to achieve very high $S$ and $ZT$, which should allow then to build nanoscale devices with high efficiencies. We also note that such designs could also be able to be achieved in the near future, due to the rapid advance of the fabrication and characterization techniques of nanogaps \cite{PriNL11,NefNl14,GehNL16,Abb17} in two-dimensional materials and related systems.The delivery of edges with tailored shapes and morphologies should also allow to implement several of the proposed systems in parallel, which would then multiply the thermoelectric performance and help to scale and integrate these systems in technologically relevant devices.

The article is structured as follows: In section II, we describe the tight-binding model that we use to describe different configurations and calculate their thermoelectric properties. In section III we present the results for two different setups: a wedge/adatom or protuberance on one side and a straight edge/flat surface on the other and wedges/adatoms or protuberances on both sides. We finish with the summary and conclusions.

\begin{figure*}
	\includegraphics[width=\textwidth]{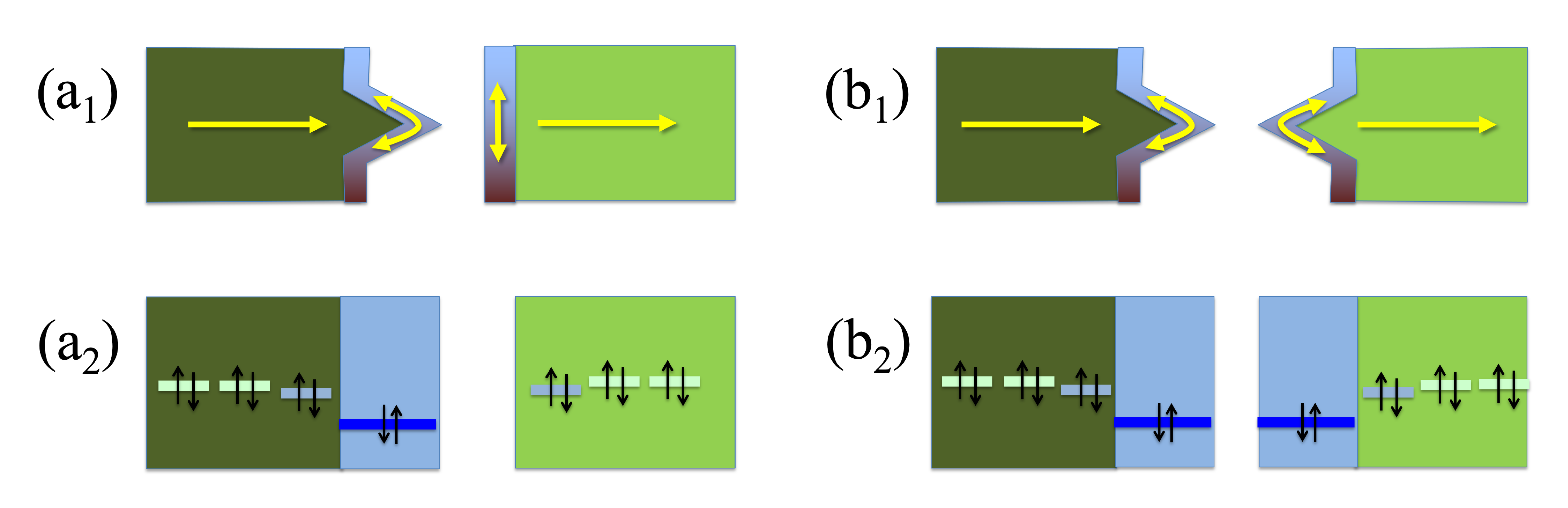}
	\caption{\label{Model} Schematic representation of the wedge-edge (a) and wedge-wedge (b) configurations. The top panels (1) show a physical representation of electrons flowing through 2D layers due to a bias voltage or a temperature gradient and the electrons on the edge states, which do not couple to the former when their group velocity is perpendicular; the bottom panels (2) show the corresponding levels of the tight-binding model.}
\end{figure*}

%%%%%%%%%%%%%%%%%%%%%%%%%%%%%%%%%%%%%%%%%%
\section{Materials and Methods}

We describe the system, which is comprised of either two facing surfaces of a bulk material or two facing edges of a 2D material, both of them separated by a gap and with or without wedges or protuberances, with a tight-binding model that takes into account the coupling between the bulk states and those states belonging to wedges/adatoms or protuberances and also the coupling between those states across the gap \cite{GarNs18,Gar20}. The model is comprised of two one-dimensional semi-infinite chains that end in a surface state and couple across a gap. We notice and stress that, even though this may seem an oversimplified system, it perfectly captures the transport properties of these or similar systems, and gives results that agree almost exactly with those calculated using more involved ab-initio simulations \cite{GarNs18,Gar20}. The model works properly because it takes into account the essential parameters that affect the electronic transport through the system. As such, it is able to accurately reproduce the transport characteristics around the Fermi level at zero bias and also at relatively large bias voltages, giving good agreements for all those regimes with the aforementioned ab-initio simulations and even with experiments \cite{GarNs18}. We also note that the model is dynamically stable, since it does not evolve over time, i.e. the sites and all parameters are kept fixed and do not change; this is relevant for systems with strong bonds whose coordinates do not change at room temperature. Therefore, it is not necessary to perform tight-binding molecular dynamics \cite{Wan96}.

The Hamiltonian includes two electrodes that couple through a vacuum gap: $\hat{ \cal H}=\hat{\cal H}_\textrm{l}+\hat{\cal H}_\textrm{r}+ \hat{\cal V}_\textrm{lr}$, where $\textrm{l}$ and $\textrm{r}$ stand for left and right electrodes, respectively. This Hamiltonian includes only electronic terms, but not  phonons because, as commented before, the transport of phonons should be negligible due to the vacuum gap between the electrodes. The most important parameters in this model are the strength of the couplings between the edge/surface states and the bulk states and the strength of the coupling between bulk or edge/surface states across the gap. The Hamiltonian of both sides (left and right), $\hat{\cal H}_\textrm{l(r)}$, is defined as follows:

\begin{equation}
	\begin{split}
		\hat{\cal H}_\textrm{l(r)}&=\sum_{\left<i,j\right>;\sigma}t_{ij\sigma}\hat c_{\textrm{l(r)},i\sigma}^\dagger\hat c_{\textrm{l(r)},j\sigma}+\sum_\sigma\epsilon_\sigma\hat d_{\textrm{l(r)},\sigma}^\dagger\hat d_{\textrm{l(r)},\sigma}+ \\
		&+\sum_\sigma t_{\textrm{1d}\sigma}(\hat c_{\textrm{l(r)},1\sigma}^\dagger\hat d_{\textrm{l(r)},\sigma}+ \mathrm{h. c.})
	\end{split}
\end{equation}

\noindent where $\hat c_{\textrm{l(r)},i\sigma}$ and $\hat d_{\textrm{l(r)},\sigma}$ represent the annihilation operators on the bulk and edge/surface states on the left (right) electrode for spin $\sigma$ ($\sigma=\uparrow,\downarrow$), respectively, $t_{ij\sigma}$ are the spin-dependent on-site energies ($t_{ii\sigma}=\epsilon_\sigma$) and coupling terms between different sites and the sum $\left<i,j\right>$ runs only to first nearest neighbours. Notice that we assume the electronic repulsion at the surface state is small enough to avoid including a large intra-atomic repulsion $U$, which means the model might not apply to strongly correlated systems with localized states and large $U$ (i.e. d or f states), where it would be necessary to go beyond a single particle description. This is in general true for surface states where electrons are delocalized between various atoms. In this case we take then a mean field approach and $U$ is included as a shift of the on-site energy level. The states at the surface or edges can have a magnetic configuration, i.e. a spin splitting $|\epsilon_\uparrow|\neq |\epsilon_\downarrow|$. In such a case, the bulk states closer to them can also be affected, which may then translate into an on-site energy slightly different from that of the other bulk states and even a small magnetic splitting ($t_{11\uparrow}\neq t_{11\downarrow}\neq\epsilon_\sigma$). We use $t_{ij\sigma}=t=-3$ eV and $\epsilon_\sigma=\epsilon= 0$ eV (the same values for both spins). To reflect the influence of the surface on the states close to it, we apply a shift to $t_{11}=\epsilon_1$ of $-0.05$ eV. We use for the surface state an on-site energy level $\epsilon=-0.5$ eV, which defines the position of the resonance in the transmission. 

The other term that enters in the Hamiltonian couples both sides of the gap and is given by

\begin{equation}
	\begin{split}
		\hat{\cal V}_\textrm{lr}&=\sum_\sigma[\gamma_{\textrm{dd}\sigma}\hat d_{\textrm{l},\sigma}^\dagger\hat d_{\textrm{r},\sigma}+\sum_{i,j=1}^2(\gamma_{\textrm{d}i\sigma}\hat d_{\textrm{l},\sigma}^\dagger\hat c_{\textrm{r},i\sigma}+\gamma_{i\textrm{d}\sigma}\hat c_{\textrm{l},i\sigma}^\dagger\hat d_{\textrm{r},\sigma}+\\
		&+\gamma_{ij\sigma}\hat c_{\textrm{l},i\sigma}^\dagger\hat c_{\textrm{r},j\sigma})+\mathrm{h. c.}].
	\end{split}
\end{equation}

\noindent This part takes then into account the couplings between bulk and/or localized states at the edge/surface across the gap. The first (second) index of the coupling elements $\gamma$ ($d$ or $i$) corresponds to a localized or bulk state in the left (right) part and vice-versa.

This last term of the Hamiltonian and the previous couplings between the bulk and surface states give rise, when computed the transmission, to features that lead to remarkable electronic and thermoelectric characteristics, as we will see later. Notice that the total Hamiltonian has also been slightly simplified with respect to the one used in previous studies, which was employed to calculate the electronic transport properties with more configurations and combinations of parameters. In this case, however, we keep only the most relevant terms that affect the thermoelectric performance. Other parameters, such as second order couplings or other couplings to the surface states that generate interference effects and which can appear in some particular cases \cite{Gar20}, are not taken into account in this Hamiltonian, because they influence mainly the transmission and the current, but have not much effect on certain thermoelectric properties (Seebeck coefficient and figure of merit). We keep then essentially the coupling between the closest states across the gap, which can be, depending on the configuration, a surface and a bulk state or two bulk states; we use the same value of the coupling in both cases, i.e. $\gamma_{\mathrm{d}1}=\gamma_\mathrm{dd}=-0.01$ eV, respectively.

The thermoelectric coefficients are obtained by calculating first the energy-dependent transmission $T(E)$ through the junction,

\begin{equation}
	T(E)=\mathrm{Tr}[\Gamma_\mathrm{L}\:G^\mathrm{R\dagger}_\mathrm{M}\:\Gamma_\mathrm{R}\:
	G^\mathrm{R}_\mathrm{M}](E)
\end{equation}

\noindent where $G^\mathrm{R}_\mathrm{M}(E)=\left[E+i\delta-H-\Sigma_\mathrm{L}^\mathrm{R}(E)-\Sigma_\mathrm{R}^\mathrm{R}(E)\right]^{-1}$ is the retarded Green's function of the scattering region, which includes part of the electrodes, $\Gamma_\alpha=i\left[\Sigma_\alpha^\mathrm{R}-\Sigma_\alpha^\mathrm{R\dagger}\right]$ and $\Sigma_\alpha^\mathrm{R}(E)=V_{\mathrm{M}\alpha}\:G^\mathrm{0R}_\alpha\:V_{\mathrm{M}\alpha}=t_{ij\sigma}^2\:G^\mathrm{0R}_\alpha$ is the corresponding selfenergy for electrode $\alpha$, with $G^\mathrm{0R}_\alpha$ the bulk retarded Green's function of the electrodes.  For this particular case, $G^\mathrm{0R}=2/(E-\epsilon+\Delta)$, where $\Delta=\mathrm{sign}(E-\epsilon)\sqrt{(E-\epsilon^2)-4\:t^2}$, the same for both electrodes. Once calculated the transmission, next step involves the evaluation  of the moments of the transmission using well known expressions that consider the transmission and the Fermi distribution function \cite{Fin09}, i.e. by extending the Landauer-B\"uttiker formalism to consider both charge and heat currents \cite{Cla96}. Notice as well that the transport is ballistic and therefore the expressions for the thermoelectric coefficients do not depend on parameters such as the electron relaxation time. The temperature dependence enters into the Fermi distribution function, or, more explicitly, into its derivative within the expressions of the moments of the transmission ($L_n(T)=\int_{-\infty}^{\infty}(E-E_\textrm{F})^nT(E)[\partial f(E,T)/\partial E]\textrm{d}E$, where $E_\textrm{F}$ is the Fermi energy and $f$ is the Fermi distribution function). The conductance, Seebeck coefficient, electronic thermal conductance and figure of merit are given in terms of such moments as

\begin{equation}
	G=\frac{2\mathrm{e}^2}{\mathrm{h}}L_0
\end{equation}

\begin{equation}
	S=-\frac{1}{\mathrm{e}T}\frac{L_1}{L_0}
\end{equation}

\begin{equation}
	\kappa=\frac{2}{\mathrm{h}T}\left(L_2-\frac{L_1^2}{L_0}\right)
\end{equation}

\begin{equation}
	ZT=\frac{S^2GT}{\kappa}=\frac{1}{\frac{L_0L_2}{L_1^2}-1}
\end{equation}

\noindent and are then evaluated at particular values of the Fermi energy and the temperature. In the figure of merit we also consider the effect of the phonon conductance, which enters as an additional factor in the denominator:

\begin{equation}
	ZT=\frac{S^2GT}{\kappa+\kappa_\textrm{ph}}
\end{equation}

\noindent We approximate such term by using a conductance per unit area of $10^7$ W/(m$^2$K), which is a typical value that can be found for distances around 3 \AA\, at room temperature \cite{Guo22}, and an area of 0.25 nm$^2$ (an square with a side of 0.5 nm, which is still relatively large for systems that have a few atoms at the tip). This gives a thermal conductance $\kappa_\textrm{ph}=2.5$ pW/K.

The model considers three possible scenarios that cover the most relevant configurations of physical nanogaps between surfaces or edges of 2D materials. These configurations, which generate a plethora of electronic effects even without any bridging component, depend on the particular structure of the edges/surfaces on both sides of the junction and produce completely different transport properties. The main difference between such configurations is due to the absence or presence in one or both sides of structural defects such as protuberances or wedges. These structural differences translate in the Hamiltonian in changes of the coupling between the bulk and the surface states ($t_{\textrm{l(r),1d}\sigma}$); this coupling is zero for straight edges/flat surfaces or has a particular finite value for wedges/adatoms or protuberances. When such coupling is not zero, it can give rise to sharp transmission resonances that signal the presence of localized states at the edges. The slope of such resonances can dramatically modify the Seebeck coefficient and, consequently, the figure of merit, following thus the strategy proposed by Mahan and Sofo to achieve the best thermoelectric performance \cite{Mah96}. We will focus then on cases that give rise to such resonances, i.e. those with wedges/adatoms on one one or both sides of the gap.

%%%%%%%%%%%%%%%%%%%%%%%%%%%%%%%%%%%%%%%%%%
\section{Results and discussion}
We have calculated results for all possible configurations that stem from different combinations of features on both sides of the gap. As commented before, these features can be wedges that protrude from edges of two-dimensional materials or adatoms/protuberances that protrude from surfaces of three-dimensional materials. The possible combination of these features with straight edges or surfaces gives three possible cases \cite{GarNs18,Gar20}. However, the simplest case, which has straight edges or clean surfaces on both sides of the gap and leads to smooth and featureless transmissions, produces rather small magnitudes of the thermoelectric coefficients and poor efficiencies, evidenced by the small figure of merit, so we will not consider it. We will focus then the other two cases, which have sharp features that can potentially lead to much higher thermoelectric coefficients: the wedge-edge (we use this notation, which refers two-dimensional materials, but the model also applies, as commented previously, to surfaces of three-dimensional materials with or without protrusions or adatoms \cite{Gar20}) and wedge-wedge configurations. Notice also that the wedge-edge configuration has the largest probability of occurring in current experimental setups and has even been measured in some recent experiments \cite{GarNs18}. We model, as commented before, non-magnetic configurations, which give a single peak in the transmission; we use in this case a surface on-site energy level $\epsilon=-0.5$ eV and an additional shift of the states close to this of $-0.05$ eV (see above).

\begin{figure*}
	\includegraphics[width=\textwidth]{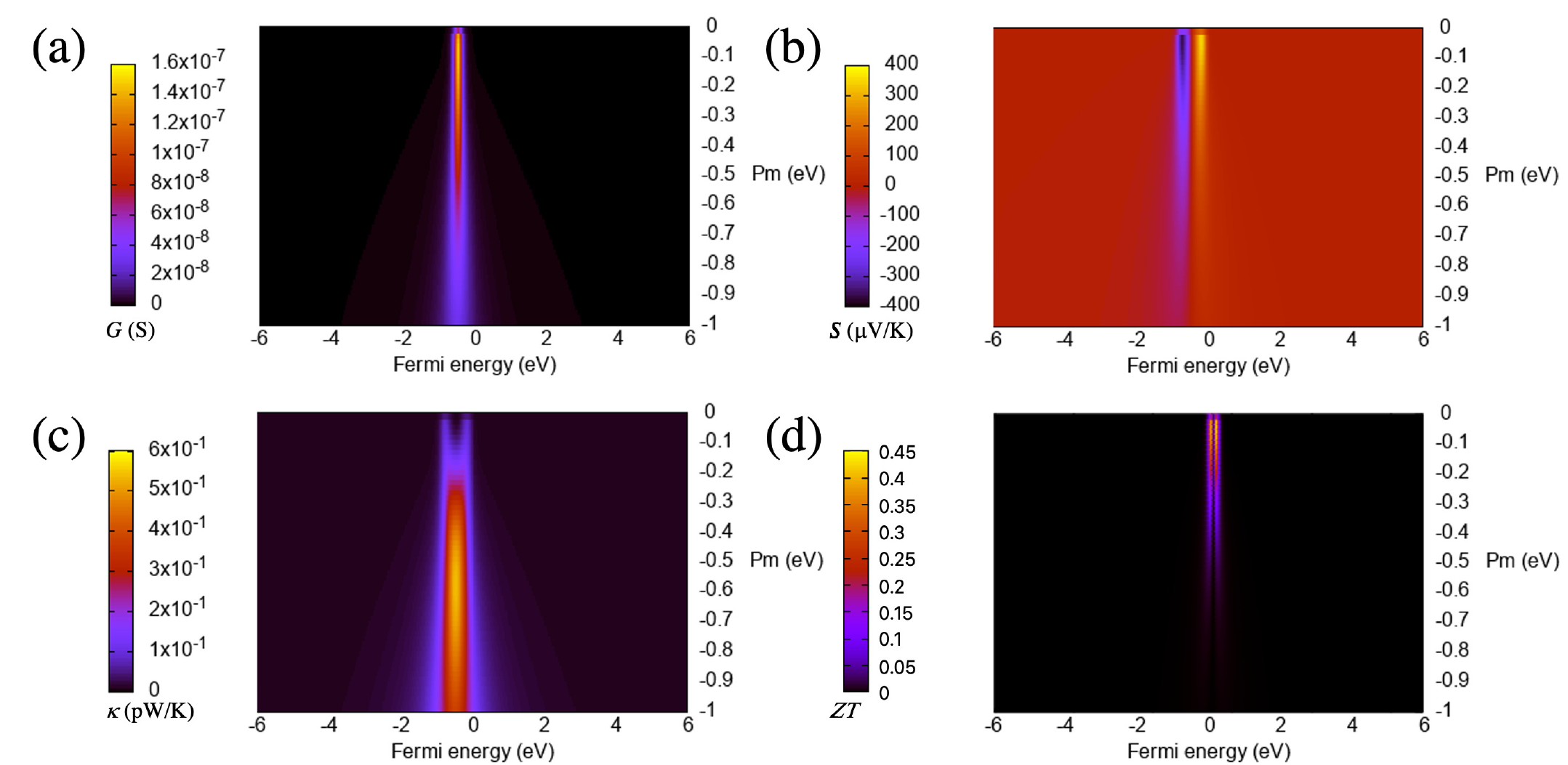}
	\caption{\label{we.NS.00.00} Electrical conductance (a), Seebeck coefficient (b), electronic thermal conductance (c) and figure of merit (d) as a function of the Fermi energy and the coupling parameter for the wedge-edge configuration. The temperature is 300 K.}
\end{figure*}

\begin{figure*}
	\includegraphics[width=\textwidth]{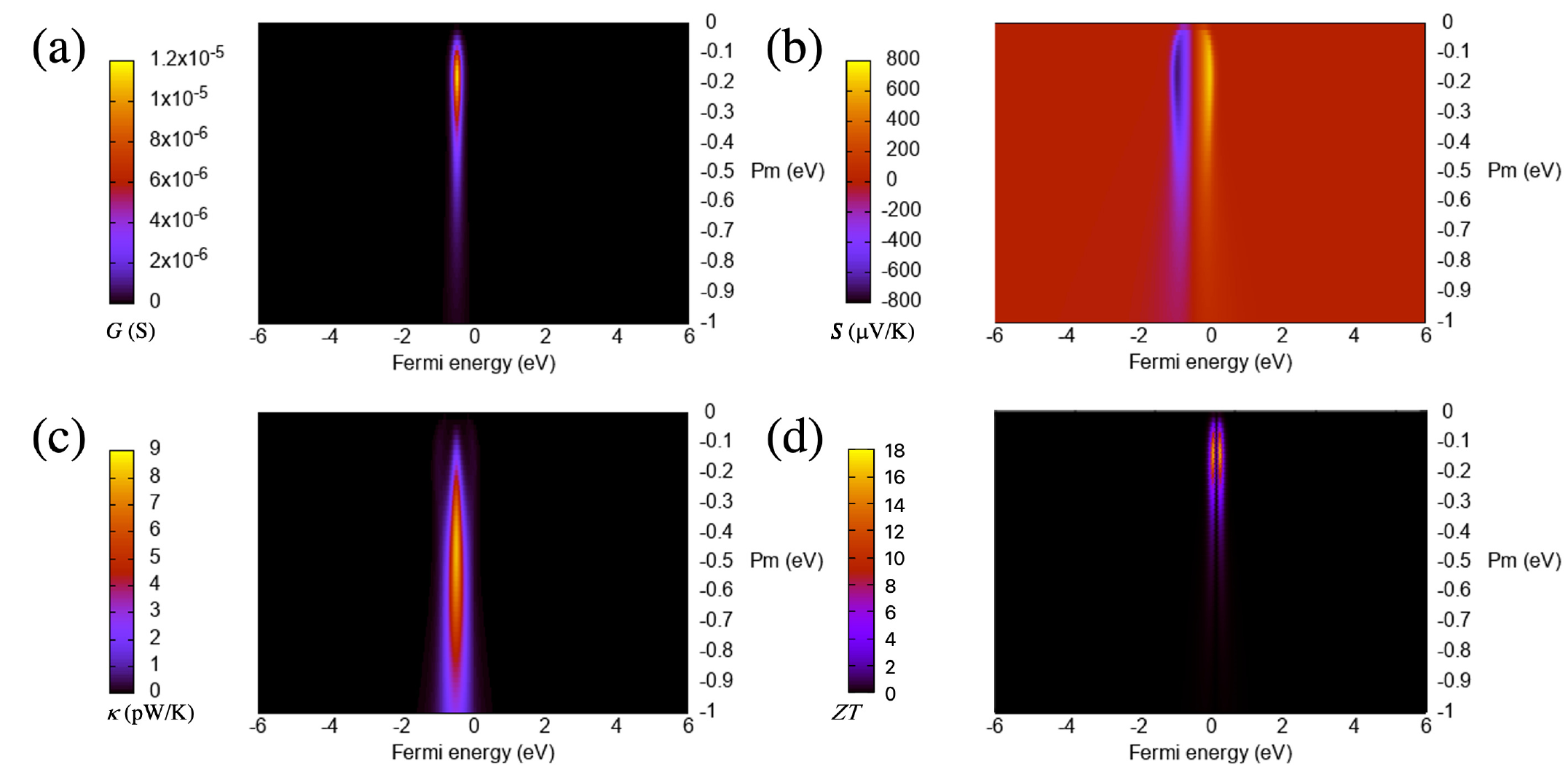}
	\caption{\label{ww.NS.00.00} Electrical conductance (a), Seebeck coefficient (b), electronic thermal conductance (c) and figure of merit (d) as a function of the Fermi energy and the coupling parameter for the wedge-wedge configuration. The temperature is 300 K.}
\end{figure*}

\subsection{Wedge-edge configuration}

The zero-bias transport properties in this case are characterized by the presence of one or two Breit-Wigner resonances in the transmission, depending on whether the system has or not spin polarization. Such resonances can be located at a given energy relative to the Fermi level and have a certain width, depending on a series of parameters that enter in the model. The position is determined by the on-site energies of the surface/edge states and the width is governed  by the coupling to the bulk states. The location and sharpness of these states are key to the generation of large Seebeck coefficients and figures of merit, since these quantities increase with the slope of the transmission at the Fermi level. 

As a function of the Fermi energy, which we assume can be smoothly changed with a gate potential, the conductance roughly follows the transmission, i.e. it has a structure with a single peak, just like a Breit-Wigner resonance. Regarding the Fermi level, notice, however, that in this case and in the following, the shifts of the Fermi level are a bit exaggerated to clearly show the shape of the coefficients. In reality, it is not possible to produce such large shifts (larger than $\sim 2$ eV), even with graphene electrodes that can be relatively close to a gate. The Seebeck coefficient changes non-monotonically and has a minimum below the on-site level ($-0.5$ eV) and a maximum above it, both of them of the same size. This particular structure comes from the derivative of the Transmission, i.e. from its slope. The thermal conductance has an structure very similar to the electrical conductance, but that can change with the couplings, as we will see later. These electrical conductances, Seebeck coefficients and thermal conductances translate into a figure of merit $ZT$ which has two equal peaks as a function of the Fermi energy. Such peaks come mainly from the Seebeck coefficient $S$, which is squared in the numerator of $ZT$ and is the main driving factor behind the thermoelectric performance, as we will see.

\subsubsection{Evolution with the coupling parameter}

The most important parameter that determines the sharpness of the resonances, which is one of the two main ingredients that, along with the position relative to the Fermi level, enhances the thermoelectric performance, is the coupling between the surface state and the bulk states. This parameter, $t_{\textrm{l(r),1d}\sigma}$, which couples the localized state to the bulk states on each side, depends on the type of material and the structural configuration (size and shape of the wedge/protusion; for instance, smaller or wider wedges in graphene give rise smaller couplings between the edge states and the bulk states) \cite{Gar20}. Although in reality it cannot be changed continuously, the study of its evolution can help to determine the type of material and features that would be needed to generate the highest thermoelectric performances.  This coupling parameter increases in general as the feature on the surface/edge (wedge/protrusion) gets sharper, since enhances the coupling between the edge state and the bulk states. The increase of the coupling, however, translates into an increase of the resonance width and therefore into a decrease of the thermoelectric performance. But the picture is not that simple and some details have to be taken into account, as we shall see.

\begin{figure*}
	\includegraphics[width=\textwidth]{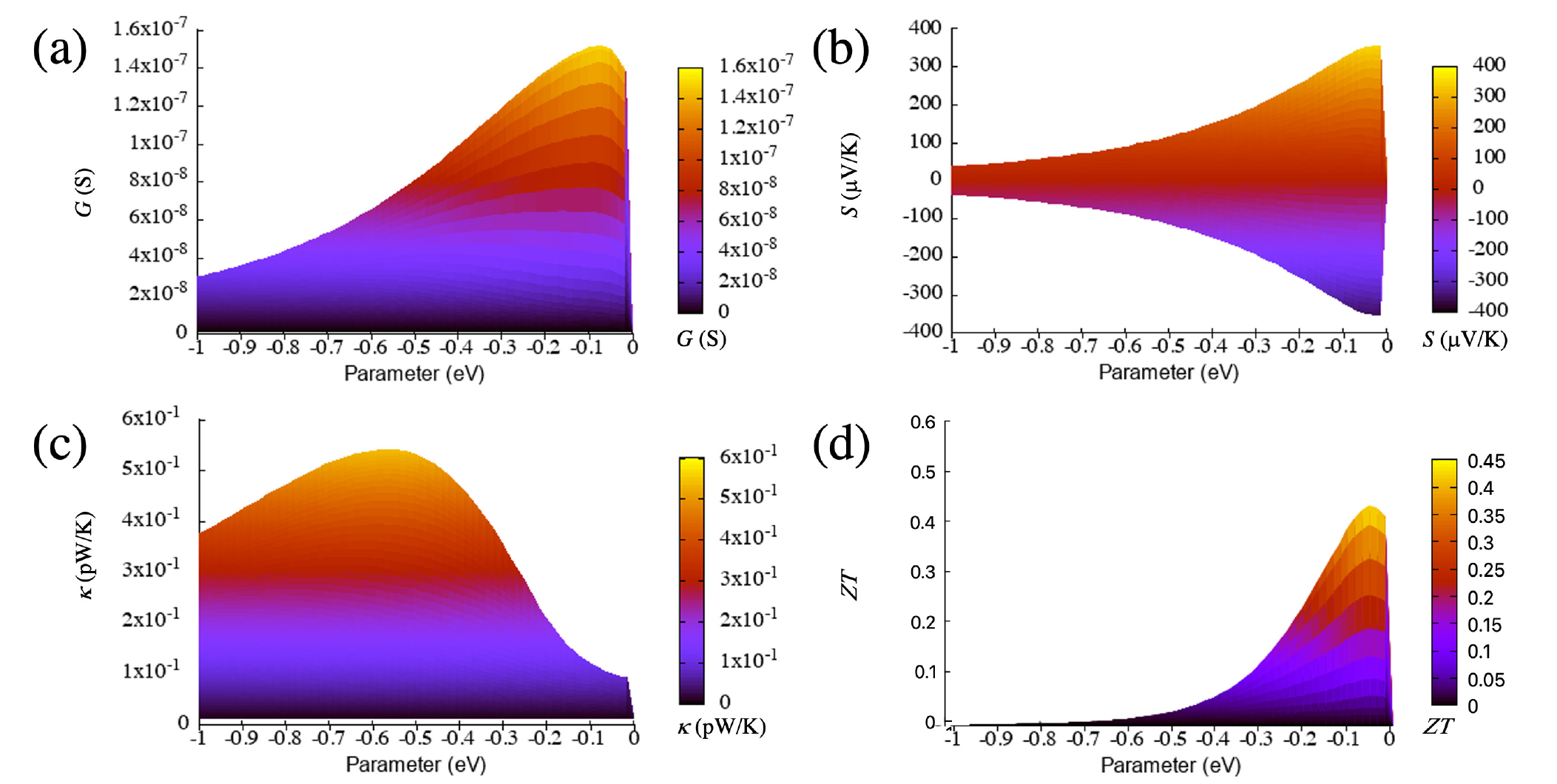}
	\caption{\label{we.NS.90.90} Electrical conductance (a), Seebeck coefficient (b), electronic thermal conductance (c) and figure of merit (d) as a function of the coupling parameter for the wedge-edge configuration (side view of Fig. (\ref{we.NS.00.00})). The temperature is 300 K.}
\end{figure*}

\begin{figure*}
	\includegraphics[width=\textwidth]{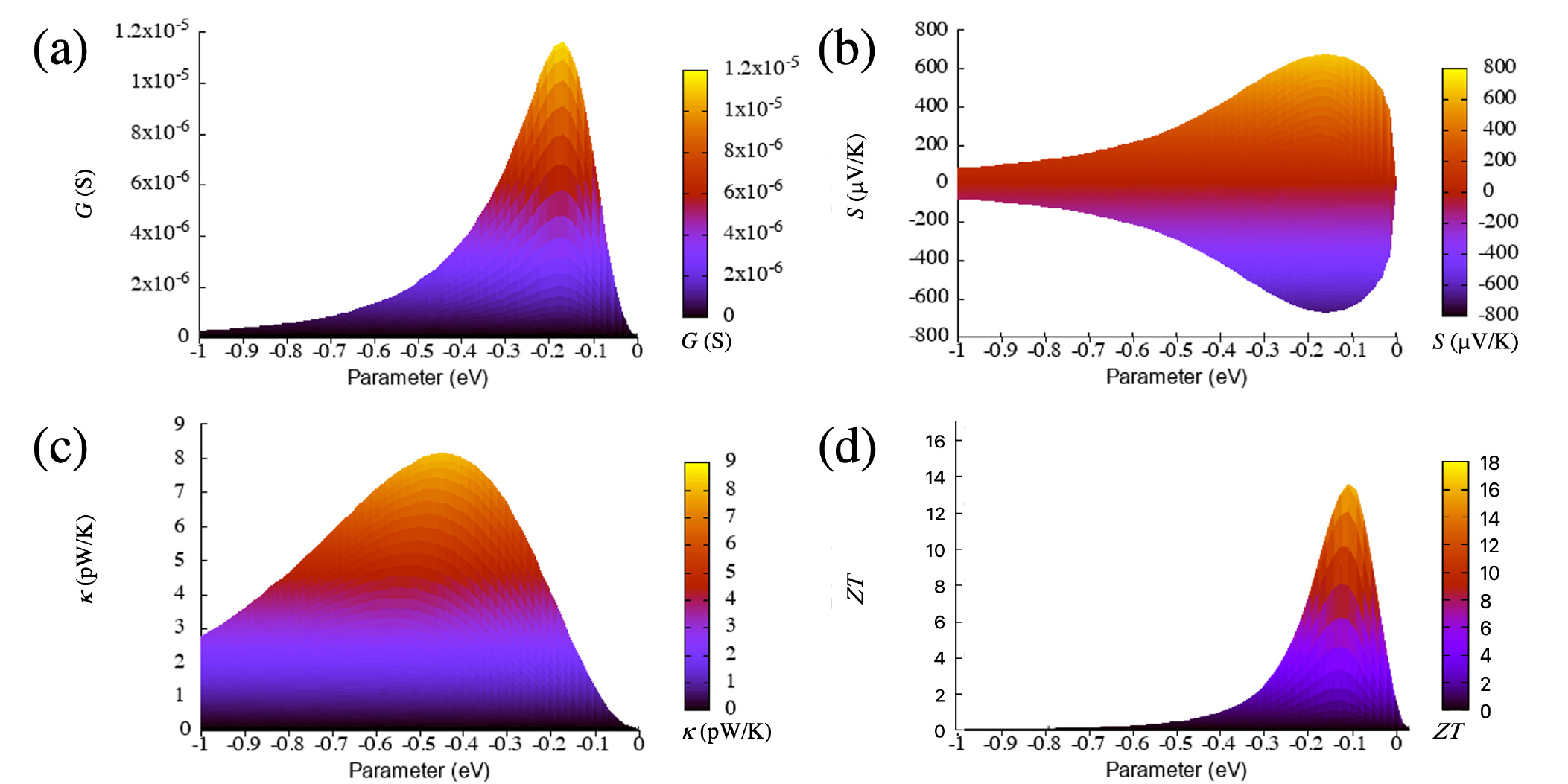}
	\caption{\label{ww.NS.90.90} Electrical conductance (a), Seebeck coefficient (b), electronic thermal conductance (c) and figure of merit (d) as a function of the coupling parameter for the wedge-wedge configuration (side view of Fig. (\ref{ww.NS.00.00})). The temperature is 300 K.}
\end{figure*}

From the curves of Figs. (\ref{we.NS.00.00}) and (\ref{we.NS.90.90}), which are calculated at room temperature (300 K),  it is clear that increasing the absolute value of the coupling parameter sharply increases the maximum of the conductance at the beginning, but then steadily reduces it, with a slope that decreases for larger values. This means that such maximum does not saturate nor remains constant, but decreases as the coupling increases after an absolute value of around 0.1 eV. This behavior of the maximum of the zero-bias conductance with the coupling parameter can be reproduced by using an analytical model for the transmission of a system with one or two resonances on each side of the gap \cite{Mar09}:

\begin{equation}\label{temar09}
	T(E)=\frac{4\Gamma_\textrm{L}\Gamma_\textrm{R}\gamma^2}{(E_{\epsilon_1}E_{\epsilon_2}-\Gamma_\textrm{L}
		\Gamma_\textrm{R}-\gamma^2)^2+(E_{\epsilon_1}\Gamma_\textrm{R}+E_{\epsilon_2}\Gamma_\textrm{L})^2},
\end{equation}

\noindent where $E_{\epsilon_{1(2)}}=E-\epsilon_{1(2)}$, $\epsilon_{1(2)}$ is the on-site energy of the level that couples to the left (right) electrodes, $\gamma$ is the coupling of the levels across the gap and $\Gamma_\textrm{L(R)}=t^2_{L(R)}\rho_\textrm{L(R)}$, with $t_{L(R)}$ the coupling between the levels and the corresponding lead and $\rho_\textrm{L(R)}$ the density of states in each lead (we assume here a wide band gap approximation, so this term is constant). In this case the coupling to the right lead ($t_\mathrm{R}$) is chosen to be smaller than that to the left lead, since for this configuration the right level is weakly coupled to the bulk states. Notice as well that the conductance here is taken as the transmission at the Fermi level; in particular, when the Fermi energy coincides with the on-site energy level, i.e. by evaluating equation (\ref{temar09}) a the maximum ($E=\epsilon_1$) and taking also into account that in these cases $\epsilon_1=\epsilon_2$, gives $T(\epsilon_1)=4\Gamma_\textrm{L}\Gamma_\textrm{R}\gamma^2/(-\Gamma_\textrm{L}\Gamma_\textrm{R}-\gamma^2)^2$ and produces a curve that, as a function of the coupling parameter $\gamma$, correctly fits those shown in panels (a) of Figs. (\ref{we.NS.90.90}) and (\ref{ww.NS.90.90}). The maximum of such curve as a function of $\gamma$ is located at $\gamma=-\sqrt{\Gamma_\textrm{L}\Gamma_\textrm{R}}$ and can therefore be changed by modifying the relative size of the couplings to the electrodes. Note that the main difference between this and next case (wedge-wedge) is the size of the coupling to the right electrode ($t_\mathrm{R}$), which, as commented before, is smaller for this configuration. According also to this model, the more asymmetric the couplings to the electrodes, the steeper the increase of the  conductance curve at small $\gamma$, in agreement with what is shown in Figs.  (\ref{we.NS.90.90}) and (\ref{ww.NS.90.90}). Such changes in the coupling give then rise not only to different conductances but also to different thermoelectric properties (both quantitatively and qualitatively) and can therefore be considered key factors that substantially affect the thermoelectric performance. Note as well that these results are general can and apply to many other nanoscale junctions which have similar resonances in the zero-bias transmission, such as those with surface states or with frontier orbitals near the contacts between the nanoscale element and the electrodes (for instance molecular electronics systems or other nanoelectronic systems with different elements bridging the electrodes).

The increase of the coupling (in absolute terms) also enhances the magnitude of the Seebeck coefficient at the beginning, as can be seen in Fig. (\ref{we.NS.90.90}). However, for larger couplings this quantity decreases due to the reduction of the slope of the transmission at the Fermi level (i.e. the resonance becomes wider and smoother when the coupling increases). This behavior can also be captured to a great extent by using as a starting point formula (\ref{temar09}) and calculating the expression for the Seebeck coefficient in terms of the transmission, which involves taking its derivative and dividing by the transmission\cite{Fin09}. This can also be easily verified in the zero-temperature limit by evaluating the expression at the Fermi energy, which gives a dependence on the derivative of the transmission (slope) divided by the transmission at that level. At finite temperatures, however, the variation of the Seebeck coefficient with the coupling deviates slightly from this behavior, because the integration in energy that gives the momenta of the transmission decreases the value of the Seebeck coefficient for very small couplings, i.e. for very sharp resonances, and masks large Seebeck coefficients in such a limit. All evolutions with temperature are anyhow very similar up to very large values and the qualitative trend is kept. In general. the smaller the temperature the sharper the features (peaks, valleys) that develop in the thermoelectric coefficients, but the smaller, however, the values of the maxima and minima that such coefficients reach, as we will see in next subsection.

The thermal conductance has two equally and clearly separated maxima for small (in absolute terms) couplings, which tend to merge as the magnitude of the coupling increases, as can be seen in Fig. (\ref{we.NS.00.00}). This is behaviour not observed for the electrical conductance, whose shape usually follows that of the thermal conductance in many cases. The reason behind this tendency is the presence of the first moment of the transmission in the expression of the thermal conductance, which is a term that has for small couplings two peaks and therefore gives rise to a thermal conductance evolution with a similar pattern.

The figure of merit, which can also be seen in Fig. (\ref{we.NS.90.90}), sharply increases first as a function of the absolute value of the coupling, but then decreases steadily for larger values. The drop is a consequence of the increase of the thermal conductance first (from 0 to $-0.5$ eV), and the decrease of the electrical conductance and the Seebeck coefficient for values smaller than  approximately $-0.05$ eV. This evolution gives rise to a maximum of $ZT$ close to 0 eV. The values at or near the maximum are rather high (around 40), which lead to a huge thermoelectric performance. Such performance is then reached, as can be seen, for a particular value of the coupling, which in this case is rather small (in absolute terms).

\subsubsection{Evolution with temperature}

\begin{figure*}
	\includegraphics[width=\textwidth]{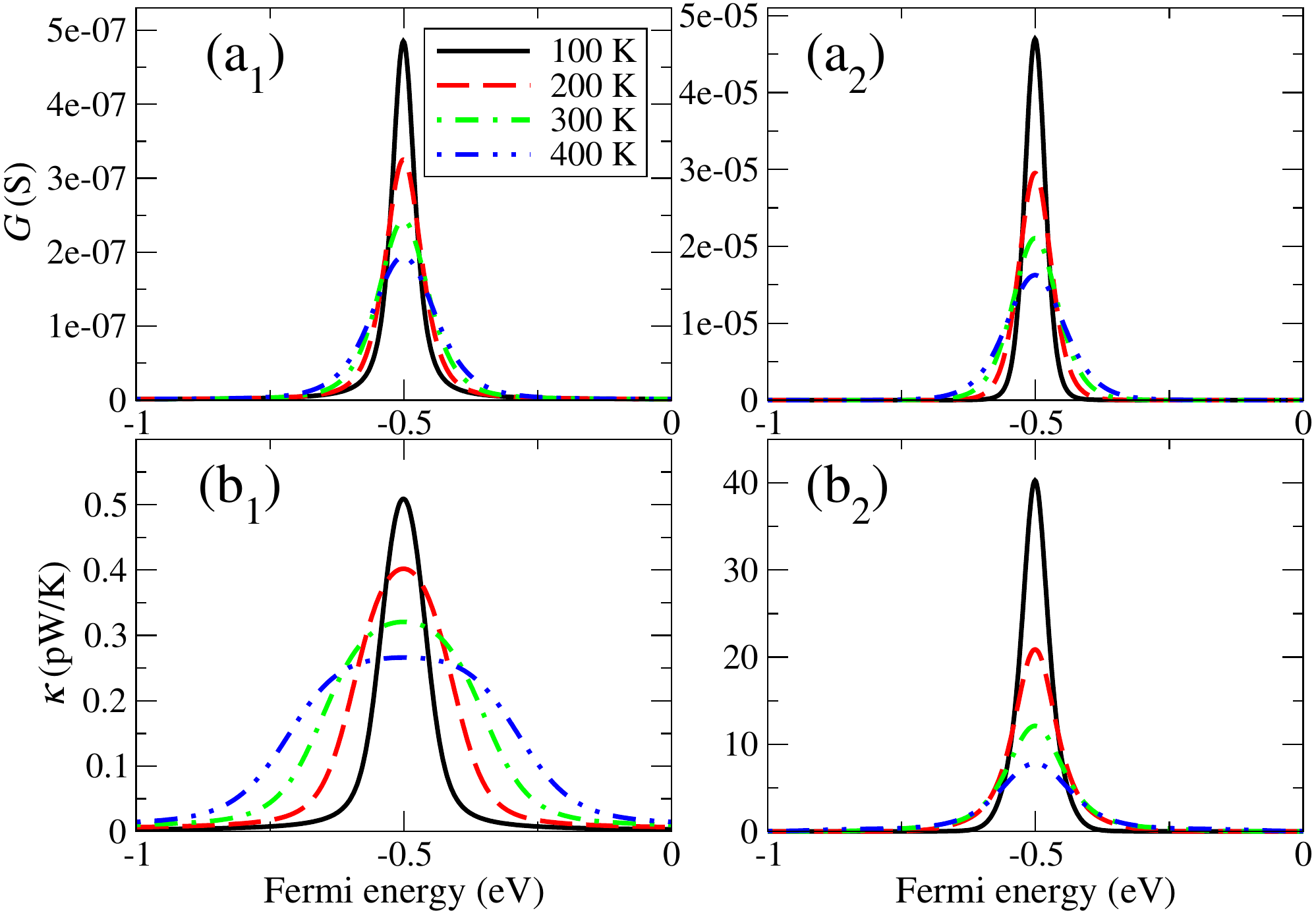}
	\caption{\label{T-dep.GK.0.20} Temperature dependence of the electronic (a) and thermal (b) conductances for the wedge-edge (1) and wedge-wedge (2) configurations. The coupling parameter in this case is $-0.2$ eV. Notice that the vertical axis intervals in panels (1) and (2) are different.}
\end{figure*}

\begin{figure*}
	\includegraphics[width=\textwidth]{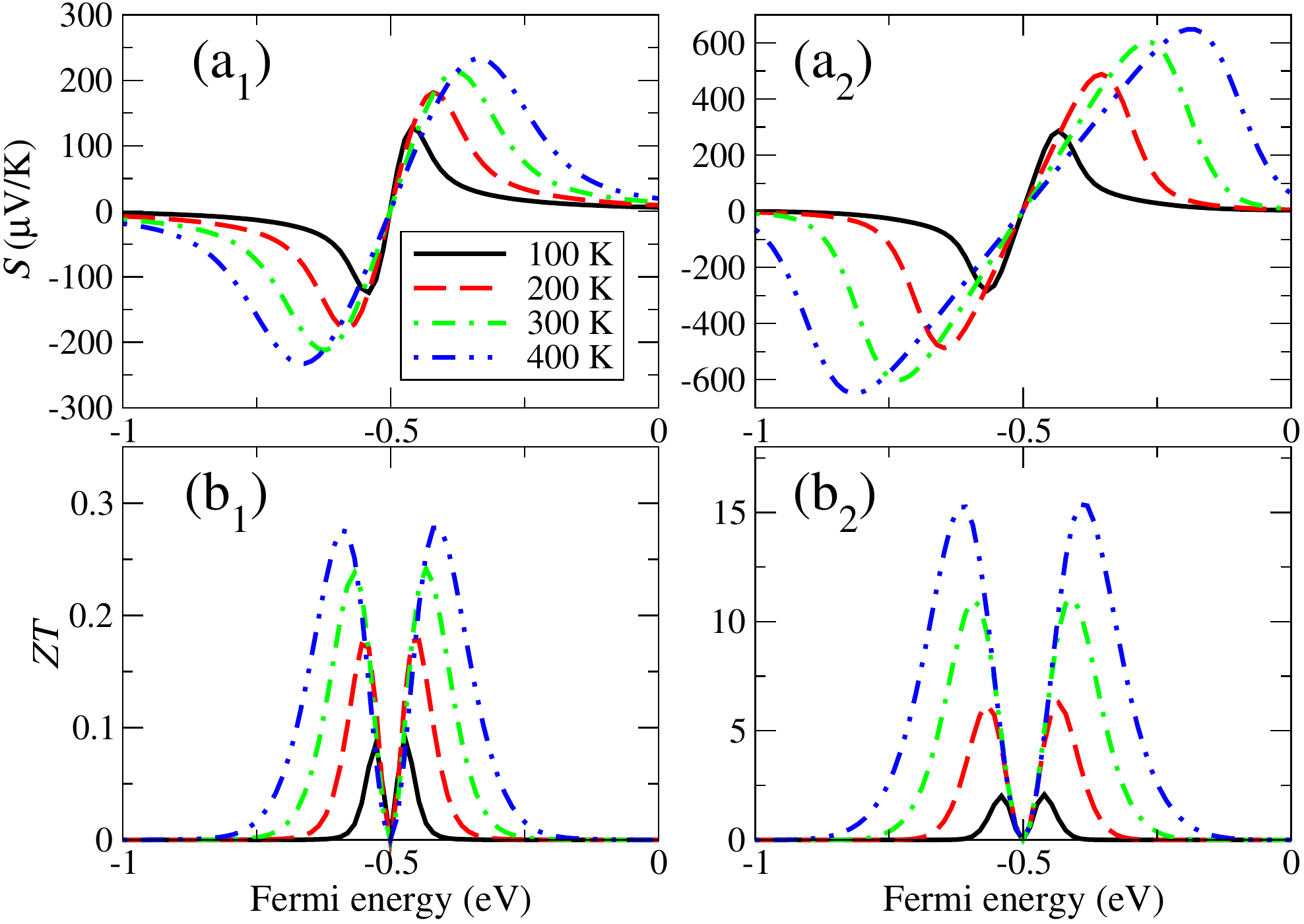}
	\caption{\label{T-dep.SZ.0.20} Temperature dependence of the Seebeck coefficient (a) and figure of merit (b) for the wedge-edge (1) and wedge-wedge (2) configurations. The coupling parameter in this case is $-0.2$ eV. Notice that the vertical axis intervals in panels (1) and (2) are different.}
\end{figure*}

Another important factor that influences the thermoelectric performance and can improve or worsen the response, is the temperature. The effect of this parameter, which enters into the Fermi distribution function, is known and has also been considered in past studies (see e.g. \cite{GarPRB13-2}), but its influence on the thermoelectric coefficients in the configurations considered in this study and the differences between them are not obvious and show very interesting trends. We have then also considered the evolution with temperature of all thermoelectric quantities. In particular, we have calculated and shown in Figs. (\ref{T-dep.GK.0.20}) and (\ref{T-dep.SZ.0.20}) the change of the thermoelectric coefficients, grouping those that look similar (electronic and thermal conductances in Fig. (\ref{T-dep.GK.0.20})), and those that show more different trends (the Seebeck coefficient and the figure of merit in Fig. (\ref{T-dep.SZ.0.20})), in a range of temperatures that can be accessed experimentally or can be meaningful for technological applications. Notice that these quantities are expected to clearly depend on temperature, since the transmission features around the Fermi level certainly feel the effect of the Fermi distribution function when the thermoelectric coefficients are calculated.

We found that the peaks of both the electrical and thermal conductances as a function of the Fermi energy widen and decrease as a function of temperature, although such trends are a bit more pronounced in case of the electrical conductance. These evolutions can be easily understood, at least in case of the electrical conductance, by taking into account the widening of the resonance given by the derivative of the Fermi distribution function that enters in the calculation for such quantities. The evolution can be seen in Fig. (\ref{T-dep.GK.0.20}). The peak of the thermal conductance, however, for high enough temperatures widens substantially and can even split in two peaks for small couplings (smaller or equal than $-0.05$ eV). This quantity can also reverse the previous trend and starts to grow slowly for larger temperatures (again, for smaller couplings, now shown in Fig. (\ref{T-dep.GK.0.20})).

The Seebeck coefficient and the figure of merit follow similar but not quite the same trends as the previous two quantities. Their evolution is shown in Fig. (\ref{T-dep.SZ.0.20}), where it can be seen that, as a function of temperature, the maxima and minima of the Seebeck coefficient widen, separate and slightly grow (for smaller couplings, such as $-0.05$ eV the height is even ketp roughly constant). The maxima of $ZT$ also widen and separate, but their height clearly increases as a function of temperature. The main responsible for such evolution is the Seebeck coefficient, whose magnitude is enhanced by the square in the expression of $ZT$, while $G$ and $\kappa$ follow similar trends and due to Wiedemann-Franz law give a contribution that is roughly proportional to $1/T$ (which cancels the $T$ in the numerator or $ZT$).

\subsection{Wedge-wedge configuration}

This configuration consists of two protuberances or wedges facing each other across the gap. Such a setup gives rise to a transmission that might look similar to that of the previous case, i.e. it has again a series of resonances (one or two in case of spin splitting) generated by localized states. However, these resonances are not the same as those from the previous case and lead to different trends in the thermoelectric coefficients, as we shall see. Note again that we consider here only non-magnetic configurations, which should allow to distinguish more clearly both types of structural configurations and univocally assign to each of them a type of evolution of the thermoelectric coefficients. We will only take then into account for this configuration, like for the previous one, a single resonance in the transmission whose shape or position does not depend on any magnetic configuration of the electrodes \cite{Gar20}.

The main differences with the previous configuration are due to the height and sharpness of the peak of the transmission, i.e. the resonance is higher and more pronounced (thinner) in this case. This is due to the symmetric nature of this configuration, because here there are localized states at the same energy on both sides of the gap and the probability of transmission is higher than in that situation where there is only one state on one of the sides. This configuration is then symmetric, although it does not give rise to Breit-Wigner resonances of height equal to one because both states are coupled asymmetrically to each side of the gap. As a function of the Fermi energy the thermoelectric coefficients are similar but not quite the same to the previous case, i.e. they have a similar shape (the conductance has a single peak, the Seebeck coefficient a maximum and a minimum above and below the on-site energy, respectively, the thermal conductance a single peak and the figure of merit two peaks around the on-site energy) but different heights and widths. The thermoelectric coefficients also show a different evolution with the coupling parameter and the temperature, as we will see, which will allow to distinguish this configuration from the previous one.

\subsubsection{Evolution with the coupling parameter}

The main parameter that influences the thermoelectric properties is again the coupling between the surface states  and the bulk states, $t_{\textrm{l(r),1d}\sigma}$ (which in this case appears on both sides in this configuration). We assume the junction is symmetric, i.e. the type of material and structural configuration (size and shape of the wedge or protuberance) are the same on both sides and therefore the coupling parameters are equal. We then evolve equally the coupling parameters on both sides to study the behavior of the thermoelectric coefficients. The evolution of most of the thermoelectric quantities and, specially, the Seebeck coefficient and the figure of merit, with this parameter is again not evident, as can be seen in Figs. (\ref{ww.NS.00.00}) and (\ref{ww.NS.90.90}), respectively, where we plot such quantities as function of this parameter and energy. Both of these quantities have rather high values (in absolute terms), which are clearly higher than those in the previous case. Also, the evolution, although similar in some ranges, is not exactly the same as that of the previous configuration and can be used to distinguish both configurations. We describe in detail the characteristics of such evolution and the differences with the previous configuration in the following.

We focus first again on the electrical conductance, whose evolution can also be captured by taking as reference equation (\ref{temar09}) and deriving the thermoelectric coefficients from it. On one hand, the calculated conductance in this configuration steadily increases from zero, like in the wedge-edge case, but keeps growing a bit more until moderate magnitudes of the coupling ($\sim -0.1$ eV), as can be seen in Fig. (\ref{ww.NS.90.90}). This means that the maximum moves to larger absolute values of the coupling. The behaviour for large magnitudes of the couplings is also different from that of the wedge-edge configuration, where the conductance decreased slowly with it. In this case the decrease is more pronounced. On the other hand, the values of this quantity are much higher than those in the previous case, almost two orders of magnitude larger. This shows already that using then a simple transport property such as the electrical conductance, its evolution with some parameters and its magnitude, it is possible to distinguish and characterize different nanogap configurations.

The absolute value of the Seebeck coefficient has also two maxima (i.e. it has a maximum and a minimum above and below the on-site energy, respectively) like in the wedge-edge configuration. This parameter steadily grows as the coupling increases from zero, but then slowly decreases, as can be seen in Fig. (\ref{ww.NS.90.90}). The maxima move also to larger absolute values of the coupling and the values are two times larger than in the previous configuration. This coefficient provides then another example of quantity that can be used to distinguish both types of nanogap configurations and shows again that, although the transmissions are rather similar (one or various peaks), the quantities derived from it behave differently as a function of certain parameters. This proves as well that those quantities can be very sensitive to very small changes derived from different structures or compositions on both sides of the gap. 

The thermal conductance has a single peak for most of the range of te coupling parameter (see Fig. (\ref{ww.NS.00.00})), as opposed to the previous case, which had two peaks for a certain range (small couplings) and a wider structure. The growth and decline of this quantity is more pronounced than in the wedge-edge case but somehow similar, as can be seen in Fig. (\ref{ww.NS.90.90}). The maximum, however, is a bit more pronounced and moves slightly to smaller absolute values of the coupling. This evolution is due again to the different magnitude and shape of the resonance, which produces also larger absolute values of this quantity (roughly one order of magnitude larger than in the previous case).

The figure of merit, on the other hand, has a clear maximum as a function of the coupling parameter and increases initially more smoothly than in the wedge-edge configuration, as can be seen in Fig. (\ref{ww.NS.90.90}). This behaviour is again due to the dependence of the figure of merit on other variables such as the electronic and thermal conductances and the Seebeck coefficient. The first two quantities increase first with the coupling, but the increase is however more pronounced for the thermal conductance; this, along with the small decrease of the Seebeck coefficient, gives rise to the maximum of the figure of merit as a function of the coupling parameter, after which this quantity decreases again rather sharply (but not as much as in the previous case). The maximum moves as well to larger absolute values of the coupling, like in the electrical conductance and the Seebeck coefficient.  Note also that in this case the magnitude of the figure of merit is much higher than in the wedge-edge configuration, with values larger than 200. This implies that higher thermoelectric efficiencies can be achieved with symmetric configurations that lead to higher resonances in the transmission.

\subsubsection{Evolution with temperature}

The evolution with temperature of the thermoelectric coefficients is also shown in Figs. (\ref{T-dep.GK.0.20}) and (\ref{T-dep.SZ.0.20}). This evolution is similar to that of the wedge-edge configuration, but the exact dependencies of some coefficients are different. In case of the  electrical conductance, its evolution is very similar and almost indistinguishable to that of the previous configuration. However, the total magnitude of this quantity is much higher (almost two orders of magnitude). The thermal conductance, however, decreases more steadily and has a maximum that does not widen too much as the temperature increases, keeping in this case an evolution that looks similar to that of the electrical conductance. This shows that the more symmetric the coupling, the more similar the evolution of both quantities.

In case of the Seebeck coefficient, the increase of the height and the separation of the maxima/minima are more pronounced than in the previous configuration and the absolute values are larger as well. This is a consequence of the less asymmetric configuration, produced by the presence of two localized states with the same energies that couple across the gap. This symmetry increases the magnitude of the peak and therefore the range where the derivative is high. The same behavior can also be seen for the figure of merit, whose peaks have clearly a more pronounced increase than in the previous configuration and reach higher values. This shows  that the more symmetric the configuration in these systems, the higher the thermoelectric performance.

%%%%%%%%%%%%%%%%%%%%%%%%%%%%%%%%%%%%%%%%%%
\section{Conclusions}

We have thoroughly characterized the thermoelectric properties of nanoscale junctions based on physical gaps between surfaces or edges of two-dimensional materials. We have found that, depending on the type of feature present on one or both sides of the gap, the thermoelectric coefficients can be rather different and have distinct evolutions with certain parameters. In particular, we have found that all coefficients show a non-trivial behavior as a function of the coupling parameter between the localized states and the bulk states of the electrodes. The electrical and thermal conductances, the figure of merit and the absolute value of the Seebeck coefficient increase first and then decrease as a function of the absolute value of the coupling parameter. These increases and decreases depend on the type of configuration: for the wedge-edge configuration the increase of the electrical conductance, the Seebeck coefficient and the figure of merit is sharper for small couplings, while for the thermal conductance it is smoother; the decrease of all these quantities but the figure of merit is however smoother in this case. The position of the maxima depends also of the configuration, i.e. for the wedge-edge case all these quantities but the thermal conductance have maxima located at smaller values of the coupling. Regarding the total magnitude of the coefficients, it is much higher in the wedge-wedge case for all of them. Such evolutions can be qualitatively explained and characterized by using a simple model of an asymmetric Breit-Wigner resonance.

We have also taken into account the evolution with temperature of the thermoelectric coefficients. This parameter, which enters into the Fermi distribution function, has a sizeable influence on the absolute magnitude of these quantities. In particular, we have found a dissimilar evolution with temperature: increasing the temperature substantially decreases the magnitude of the electrical and thermal conductances, but increases however the magnitude of the Seebeck coefficient and the figure of merit. We can conclude then that the higher the temperature, the better the thermoelectric performance. Regarding the differences between configurations, the increase or decrease of all coefficients with temperature is larger and more pronounced in the wedge-wedge case.

The results also show that it is possible to distinguish structural configurations (wedge-edge or wedge-wedge) of a given gap with the thermoelectric coefficients. Although the evolutions in both cases are qualitatively similar, there are clear quantitative differences in the magnitude and in other values such as the the maxima of such evolutions. We can then conclude that symmetric configurations with localized states on both sides (wedge-wedge) give much higher thermoelectric performances, as shown by the larger values (one order of magnitude) of the figure of merit in such cases.

Finally, we note that the results presented here not only apply to surfaces or edges separated by a gap, but also to many other systems, such as those that have nanoscale objects connected between electrodes and which, in principle, can be considered as qualitatively different from those studied here. In those cases there might appear similar features in the transmission (small and sharp resonances) that come for instance from surface states or states localized near the contacts. Such features give also rise to similar transport properties (see e.g. \cite{gollum}) and lead to qualitatively similar thermoelectric results.

\appendix
\renewcommand{\thesection}{A}
\section{Nanogaps in graphene layers. Ab-initio results}

\begin{figure*}
	\includegraphics[width=\textwidth]{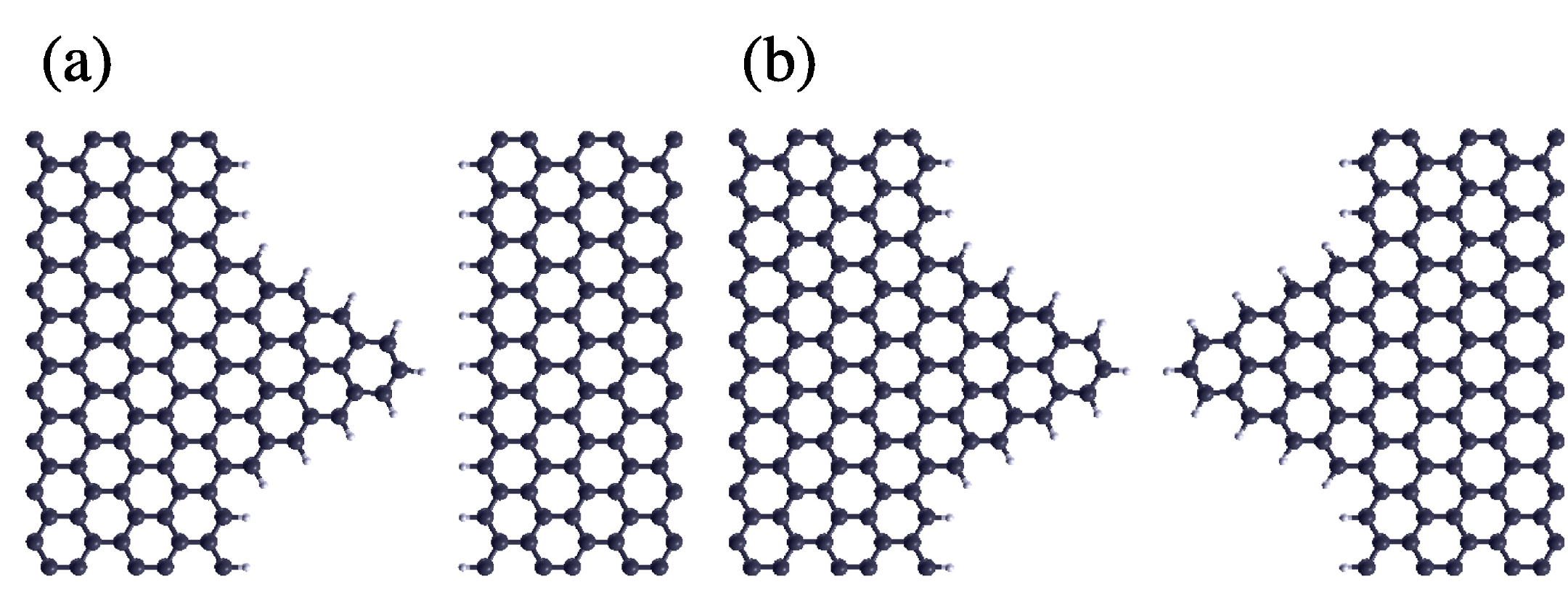}
	\caption{\label{Atoms} Atomic coordinates used to run the ab-initio simulation in the wedge-edge (a) and wedge-wedge (b) configurations. Black and white spheres represent carbon and hydrogen atoms, respectively.}
\end{figure*}

In order to compare with the results of the model, we have also simulated the thermoelectric properties of graphene layers separated by a gap. We used density functional theory \cite{Koh65}, as implemented in the Siesta code \cite{siesta}, which employs norm-conserving pseudopotentials and linear combinations of atomic orbitals. We used a double zeta (DZ) basis set, the local density approximation (LDA) \cite{pz81} and a real space grid defined with an energy cut-off  of 300 Ry. We relaxed the coordinates of the carbon and hydrogen atoms at the edges until the forces were smaller than 0.05 eV/\AA. The number of atoms, shown in Fig. (\ref{Atoms}), were 226 and  254 for the wedge-edge and wedge-wedge configurations, respectively. The size of the gap at the thinnest point, defined as the distance between the terminating  hydrogen atoms at the edges or at the tip of the wedge, was 2 \AA. Once the Siesta calculation finishes, the Hamiltonian and overlap matrices are used to calculate the thermoelectric properties with the Gollum \cite{gollum} code with a temperature of 300 K.

\begin{figure*}
	\includegraphics[width=\textwidth]{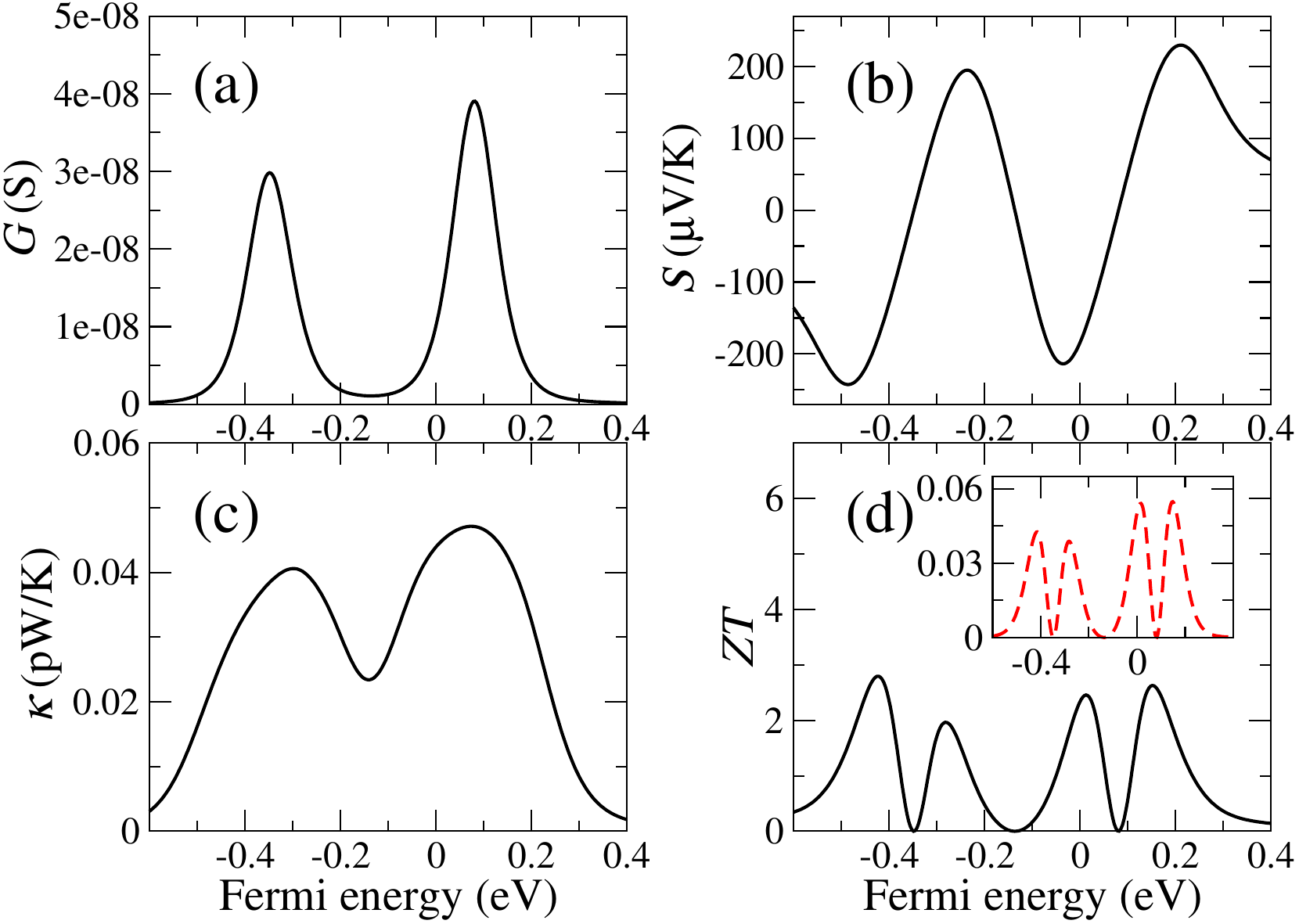}
	\caption{\label{we.Ab-initio.uu} Conductance (a), Seebeck coefficient (b), thermal conductance (c) and figure of merit (d) for a graphene wedge facing a straight edge. The inset shows the figure of merit calculated with a phonon thermal conductance of  2 pW/K.}
\end{figure*}

\begin{figure*}
	\includegraphics[width=\textwidth]{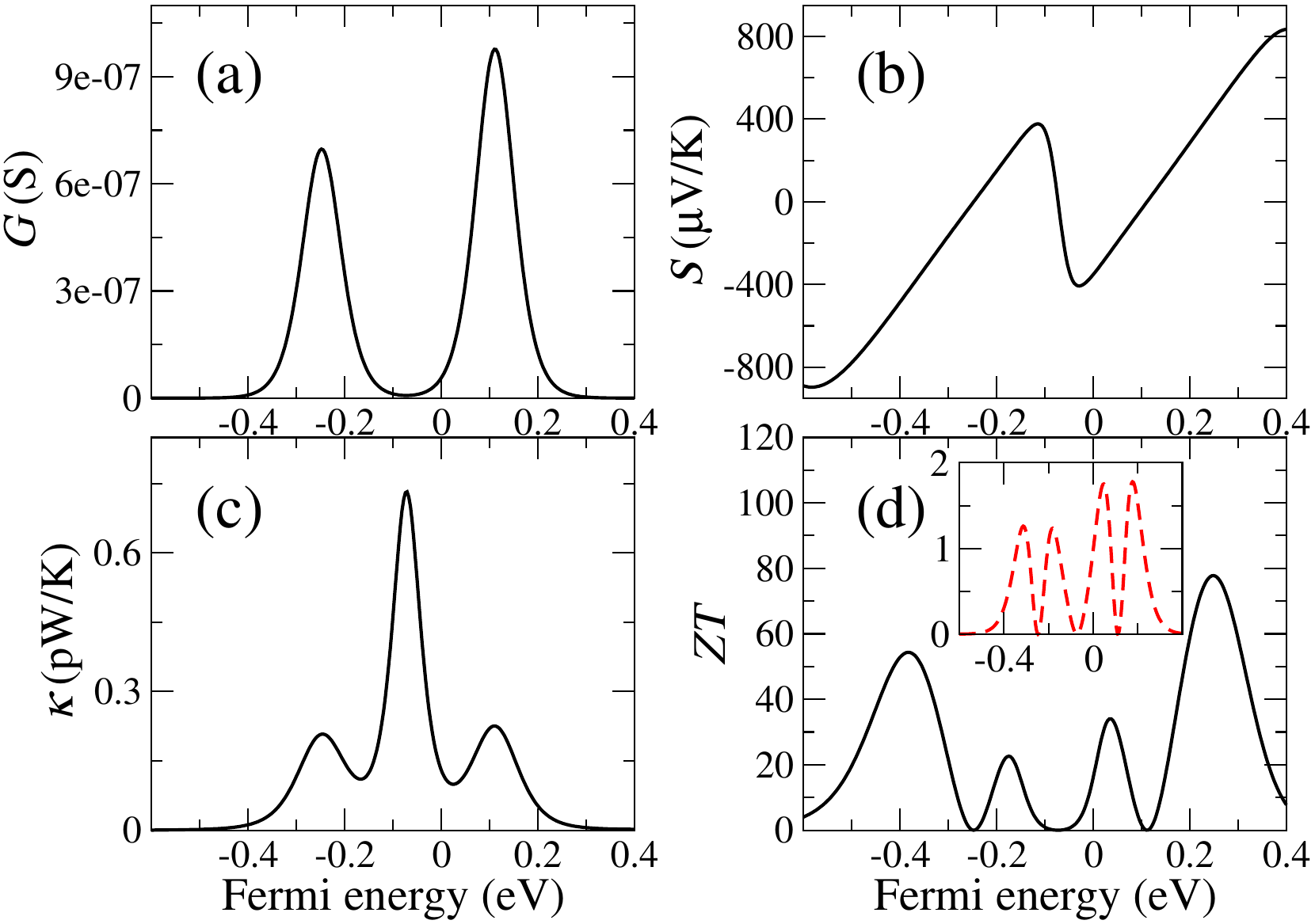}
	\caption{\label{ww.Ab-initio.uu} Conductance (a), Seebeck coefficient (b), thermal conductance (c) and figure of merit (d) for a graphene wedge facing another wedge. The inset shows the figure of merit calculated with a phonon thermal conductance of  2 pW/K.}
\end{figure*}

The phonon thermal conductance is approximated by taking a value of the thermal conductance per unit area of about $10^8$ W/(Km$^2$) \cite{Guo22} and an area of 0.02 nm$^2$ that encompasses the hydrogen atom at the tip of the wedge, which gives $\kappa_\textrm{ph}=2$ pW/K. Notice also that in this case, since the graphene edges are magnetic, the resonances are split in two. This implies that both spin components, which can have different transmissions $T^\uparrow(E)$ and $T^\downarrow(E)$, have to be used in the thermoelectric expressions: 

\begin{equation}
	G=\frac{\mathrm{e}^2}{\mathrm{h}}L_0^\mathrm{t}
\end{equation}

\begin{equation}
	S=-\frac{1}{\mathrm{e}T}\frac{L_1^\mathrm{t}}{L_0^\mathrm{t}}
\end{equation}

\begin{equation}
	\kappa=\frac{1}{\mathrm{h}T}\left(L_2^\mathrm{t}-\frac{L_1^{\mathrm{t}2}}{L_0^\mathrm{t}}\right)
\end{equation}

\begin{equation}
	ZT=\frac{1}{\frac{L_0^\mathrm{t}L_2^\mathrm{t}}{L_1^{\mathrm{t}2}}-1}
\end{equation}

\noindent where $L_i^\mathrm{t}=L_i^\uparrow+L_i^\downarrow$. The use of both spin components in such expressions gives thermoelectric coefficients that are roughly the sum of the contribution of each resonance if they are away from each other, as can be seen in Figs. (\ref{we.Ab-initio.uu}) and (\ref{ww.Ab-initio.uu}). In some cases, the resulting coefficients mix the contribution of both resonances and lead to dependencies which are not clearly split, such as the thermal conductance in Fig. (\ref{ww.Ab-initio.uu}). In any case, the results for each resonance qualitatively agree with those predicted by the model. Notice as well that, even though these particular systems based on graphene layers were not particularly tailored to produce large thermoelectric coefficients, the resulting thermoelectric performances are still good, specially those of the wedge-wedge case, which again shows the potential of these systems as future elements in thermoelectric devices.

\section*{Acknowledgements}
We thank the Spanish Ministerio de Ciencia, Innovaci\'on y Universidades for funding through the project PGC2018-094783-B-I00.

\bibliography{master}
\bibliographystyle{apsrev}

\end{document}